# Robots as-a-Service in Cloud Computing: Search and Rescue in Large-scale Disasters Case Study

Carla Mouradian*, Sami Yangui*§, Roch H. Glitho*

*CIISE, Concordia University, QC, Canada
§LAAS-CNRS, Université de Toulouse, INSA, Toulouse, France
{ca_moura,s_yangui,glitho}@encs.concordia.ca

*Abstract*— Internet of Things (IoT) is expected to enable a myriad of applications by interconnecting objects – such as sensors and robots – over the Internet. IoT applications range from healthcare to autonomous vehicles and include disaster management. Enabling these applications in cloud environments requires the design of appropriate IoT Infrastructure-as-a-Service (IoT IaaS) to ease the provisioning of the IoT objects as cloud services. This paper discusses a case study on search and rescue IoT applications in large-scale disaster scenarios. It proposes an IoT IaaS architecture that virtualizes robots (IaaS for robots) and provides them to the upstream applications as-a-Service. Node- and Network-level robots virtualization are supported. The proposed architecture meets a set of identified requirements, such as the need for a unified description model for heterogeneous robots, publication/discovery mechanism, and federation with other IaaS for robots when needed. A validating proof of concept is built and experiments are made to evaluate its performance. Lessons learned and prospective research directions are discussed.

## I. INTRODUCTION

Internet of Things (IoT) interconnects various and addressable devices over the Internet in order for them to communicate and cooperate when specific applications are implemented [1]. Radio-Frequency IDentification (RFID) tags, sensors, and actuators are among the examples of IoT devices. Similarly, IoT applications are quite diverse, ranging from aeronautics to healthcare and disaster management.

IoT systems are made up of a set of data sources (e.g., sensors) that forward the generated data to a centralized or semi-centralized collection point for analysis and processing [2]. Furthermore, the required computing resources and the used IoT devices are predefined and cannot be dynamically (un)allocated on-demand during runtime [2]. This leads to several drawbacks in terms of flexibility, cost efficiency, and scalability when IoT applications are operated. Notably, cloud computing can help address these drawbacks.

Cloud computing is a novel model for enabling ubiquitous, convenient, and on-demand network access to a shared pool of configurable computing resources [3]. It handles and delivers resources based on three predefined service models: (a) Infrastructure-as-a-Service (IaaS), (b) Platform-as-a-Service (PaaS), and (c) Software-as-a-Service (SaaS). Cloud applications are provisioned using PaaS and are offered as SaaS. The underlying required resources are provided by IaaS.

To provision IoT applications in the cloud, their lifecycle needs to be supported according to the cloud applications lifecycle setting. However, it should be noted that IoT devices have specificities and cannot be always provisioned in the same way regular cloud resources are [4]. To make IoT systems cloud-aware, the IoT context characteristics and specificities should be taken into consideration. For instance, the latency-sensitivity requirement of IoT applications needs to be considered when the applications are moved to a distant cloud accessible via the Internet [5]. The same applies to the limited computation and autonomy capabilities of IoT devices, with regards to VMs, when they are virtualized and used (e.g., see [6] for the case of wireless sensors network). Finally, the strong diversity and heterogeneity of IoT devices are often difficult to aggregate and integrate under the same ecosystem. Therefore, the publications in the relevant literature mostly focus on a specific kind of IoT applications (e.g., healthcare) and/or IoT devices (e.g., robots).

Designing IoT architectures in the cloud is challenging. On the one hand, adopting classical client-server approach may not be the best pattern. For instance, in most use cases, IoT devices are required to make some decisions locally or to communicate with other device(s) based on the collected data. On the other hand, a fully distributed pattern fails when the technical constraints and limited capabilities of IoT devices do not allow them to have heavy computing and storage resources.

This paper discusses a case study on search and rescue IoT applications in large-scale disaster scenarios. The applications are provisioned as SaaS in a dedicated PaaS and they use robots services provided by the underlying IaaS. The proposed architecture aims to enable flexible, elastic, and cost-efficient use of robots, benefiting the cloud advantages such as virtualization and scalability. The rest of the paper is organized as follows: Section II details the case study. Section III lists the related works. Sections IV and V discuss the proposed architecture. Section VI describes the validating prototype and measurements. Section VII concludes the paper and discusses the lessons learned.

## II. THE CASE STUDY: IOT APPLICATIONS WITH ROBOTS FOR LARGE-SCALE DISASTERS MANAGEMENT

This section introduces the case study at hand. It discusses the architectural challenges and requirements that would enable its provisioning in the cloud.

### A. Context

In the last decade, a myriad of novel IoT applications has emerged in large-scale disaster management. These applications aim to intervene in extreme and high-risk conditions, to for example seal a leak in a nuclear reactor or coordinate search and rescue missions when natural disasters such as earthquake occur. In most cases, these applications communicate with the on-site robots. According to ISO 8373 [7], a robot is an actuated, programmable mechanism able to perform intended tasks by moving in its environment. Robots are made of actuators and sensors and, unlike humans, can be deployed to dangerous sites. This case study concerns search and rescue application that deploys robots to disaster sites and controls them.

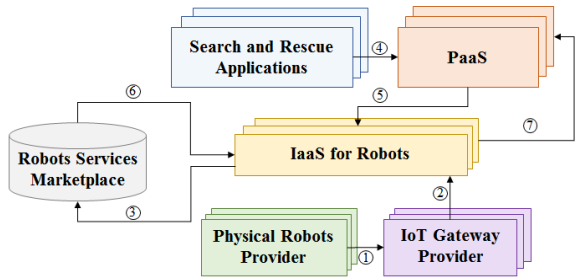

Fig. 1. Business Model

### B. Scope and Challenges

In this case study, the focus is on the IaaS aspects. Operating search and rescue applications in the cloud setting requires virtualizing the robots at IaaS and providing them to the hosting PaaS as cloud infrastructure services.

One challenge here is to perform the virtualization at two levels: node and network. The two levels of abstraction contribute flexibility and cost efficiency, which are the two key features of cloud computing. On the one hand, the robot node-level virtualization enables multiple applications to reside in a single robot and run concurrently. On the other hand, the robot network-level virtualization enables the dynamic formation of subsets of robots services, with each subset dedicated to a certain application at a given time [8].

A second challenge is the composition of robots services when needed. Some applications may need complex capabilities with two or more robots services required to perform a search and rescue task. For instance, in addition to arms that sift debris, a specific type of arms to carry fire extinguishers might be required by the same application in an earthquake site. Capabilities specific to different robots and even to different IaaSs might be needed to build the required composed services.

A third challenge is to enable the scalability of the architecture in terms of the number of robots. It should support the provisioning of a huge and various number of robots services to cover the several variants and specificities of the search and rescue applications.

### C. Requirements

Based on the identified challenges, the architecture needs to meet a set of requirements. As the first requirement, the architecture needs *a common and unified model* that enables describing the capabilities of several and heterogeneous robots, independently from their brands, technical constraints, and IaaS provider. As the second requirement, the architecture needs a *publication/discovery mechanism* for the supported capabilities of robots. The third requirement is the need for a *composition mechanism* that enables orchestrating several robots services (if needed) when none of the elementary published services satisfies the requests. The fourth requirement is the need to *federate the IaaSs for robots*. The orchestrated robots services could either belong to the same IaaS or to different IaaSs owned by different entities.

### III. STATE OF THE ART

This section lists and describes the relevant research. Although in its early days, enabling robots in the cloud has started to be the topic of a few research works. For instance, the authors in [8] propose an architecture for robotic applications as cloud computing services. The proposed architecture enables the network-level virtualization of robots and considers delegating tasks to robots belonging to other IaaSs. It supports heterogeneous robots in each IaaS. However, it does not include a model to describe these heterogeneous robots. In [9], the authors propose a cloud infrastructure that receives images from a vision acquisition system, processes the load received from the system, and accordingly controls the robots' behavior. The authors in [9] do not consider heterogeneous robots as their solution applies only to one type of robot hardware (i.e., iRobot). Consequently, neither of these works meets the first requirement concerning a unified model for robots description.

In [10], the authors present Cloud Enabled Robotics System where robots offload their computationally intensive tasks to the cloud. Robot Operating System (ROS), a robotic middleware to develop robot software, is used as the robot platform. This allows the cloud to communicate and send commands to a heterogeneous team of robots. However, the authors do not discuss how to discover and composite the robots services. So, this work does not meet the second and third requirements considering the publication/discovery mechanism and the need to composite robots services.

In [11], the authors propose a Robot *as-a-Service* platform that provides easy access to heterogeneous robots. The proposed design consists of an OCCI extension that models cloud robotics *as-a-Service*. It also includes a gateway for hosting mobile robot resources. The proposed platform allows users to have a unified view of all robots. The proposed platform's main function is to provide robot gateway and hence architectural modules that virtualize robots' capabilities and orchestrate them are not discussed. In [12], the authors propose DAvinCi PaaS. It is a software framework for data-intensive robotic cloud applications. The DAvinCi server collects data from robotic applications using ROS. Once the data is collected from the robots, it is pushed into a Hadoop system for analysis. The results are sent back to the robotic applications. The architecture enables the teams of heterogeneous robots to share data and communicate with the remote server for heavy computation. However, how these teams of robots are composed is not discussed. In [13], the authors propose Rapyuta, an open source PaaS framework for robotic applications. Rapyuta computing environment allows robots to easily access to the RobotEarth knowledge repository. The latter enables robots to benefit from the experience of other robots. Rapyuta allows robots to offload heavy computation to the cloud. It dynamically provides secure computing environments for the robots. These computing environments are tightly interconnected, allowing robots to share their services and information with other robots. However, orchestrating these robots services is not discussed. Consequently, none of these works meets the third requirement concerning a composition mechanism that enables orchestrating several robots services.

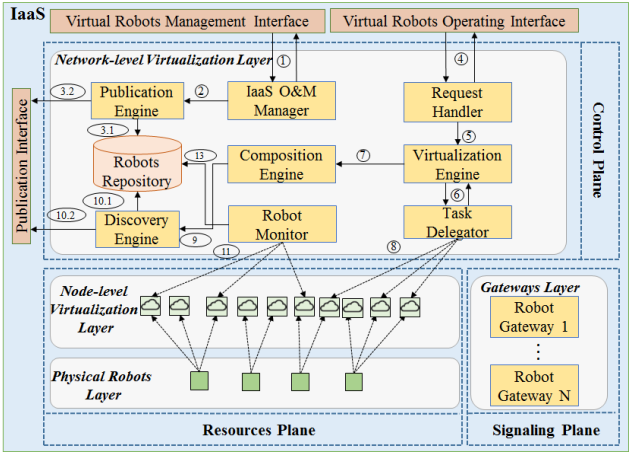

Fig. 2. IaaS for Robots Architecture

## IV. IaaS for Robots: Architecture

This section presents the case study architecture. Here, the business model is first introduced and then, the detailed architecture of the IaaS for robots is discussed. Finally, the specifications of the designed interfaces are presented.

### A. Business Model

The related business model uses and extends the *pay-as-you-go* cloud model. The robots are provisioned *as-a-Service*. The specific actors of this case study and their relations are schematized in Fig. 1. In the southbound, the *Physical Robots Providers* represent the concrete pool of the heterogeneous physical robots. The *IoT Gateway Provider* provides the required communication gateways to interact with the robots (Fig. 1, action 1). The virtualization of the robots is performed by the *IaaS for Robots Providers* (action 2). These providers publish their supported robots services in a common *Robots Services Marketplace* (action 3). The marketplace lists and indexes all the available robots for prospective use.

In the northbound, The *Search and Rescue Applications* are provisioned as SaaS over the several *PaaS*s (action 4). *PaaS*s interact with the underlying *IaaS*s to settle the required runtimes for hosting and executing these applications. They allocate the necessary robots services from the *IaaS for Robots Providers* and bound them to the applications (action 5). If the required robot service is not supported by the local *IaaS for Robots Providers*, the latter requests the *Robots Services Marketplace*

TABLE I. EXAMPLES OF THE API OPERATIONS ON THE PUBLICATION INTERFACE

| REST Resource | Operation | HTTP Action and Resource URI |
|---|---|---|
| Robot Presence Information | Create: PUBLISH presence information for newly purchased robots (joining publication) | POST: /robots |
| | Read: SUBSCRIBE for presence information of a robot | POST:/robots/{robotid}?fromuri={subscriberuri} |
| | Read: SUBSCRIBE to a list of robots | POST:/robots?fromuri={subscriberuri} |
| | Update: re-PUBLISH presence information of a robot, update an already created resource. (state change publication) | PUT:/robots/{robotid} |
| | Delete: un-SUBSCRIBE from presence info of a robot | DELETE:/robots/{robotid}/{subscribeid} |
| | Delete: un-SUBSCRIBE from presence info of list of robots | DELETE:/robots/{subscribeid} |

(action 6) to get it from another *IaaS for Robots Providers* and deliver it to *PaaS* (action 7).

### B. Detailed IaaS for Robots Architecture

The proposed IaaS architecture is shown in Fig. 2. It consists of *Resources Plane*, *Control Plane*, and *Signaling Plane*. The *Resources Plane* includes two layers: The *Physical Robots Layer* and the *Node-level Virtualization Layer*. The *Physical Resources Layer* involves the supported robots. It includes the physical heterogeneous robots with their various capabilities and characteristics. The *Node-level Virtualization Layer* contains the pool of the virtualized robots. The *Signaling Plane* contains a set of communication gateways called *Robot Gateways*. These gateways allow of hiding the heterogeneity and specificities of the robots in terms of user APIs, communication protocols, and so on. Their role is to map between the *Network-level Virtualization Layer* (in the *Control Plane*) requests and the proprietary robots APIs. The gateways are (un)instantiated on-demand in accordance with the evolution of the applications' workload and the used robots. Their design is based on our previous work described in Ref. [14].

The *Control Plane* includes the *Network-level Virtualization Layer*. The *IaaS O&M Manager* is responsible for adding the supported robots to the IaaS or removing them from it through the *Virtual Robots Management Interface* (Fig. 2, action 1). For instance, when a new robot service is added, this module parses the robot metadata (e.g., communication protocol, list of capabilities) and generates a descriptor based on a well-defined model (see Section V.A). The descriptor is then forwarded to the *Publication Engine* (action 2) that stores it locally (action 3.1) and publishes it in the remote marketplace through the *Publication Interface* (action 3.2).

Besides, the *Control Plane* exposes a *Virtual Management Operating Interface* that allows PaaS to (un)provision robots services. The front-end module is the *Request Handler*. It is responsible for analyzing the upcoming requests (action 4), and providing a set of inputs, such as task requirements, to the *Virtualization Engine* (action 5). The *Virtualization Engine* basically performs the network-level virtualization of the robots' capabilities. This is done by running an appropriate algorithm for coalition formation in multi-robot systems. The algorithm is designed and implemented as part of our previous work [15]. It is a coalition formation algorithm for large-scale disasters, ensuring the optimal coalition of robots is selected with the required capabilities for the search and rescue task. It should be noted that the *Virtualization Engine* handles local robots services but it may use robots services that belong to remote IaaSs in the coalition. This could be done through inter-IaaS cooperation to satisfy complex requests. Consequently, scheduling (action 6) and composition (action 7) techniques are often needed before delivering the final services to PaaS. On the one hand, the *Task Delegator* makes a bridge between the virtualized robots and the physical ones (action 8). It schedules the task assignment requests between the running robots services at the *Virtualization Engine* and the involved physical robots. As stated earlier, the physical robots can be either local or belong to other IaaSs. Moreover, the *Task Delegator* may also receive task assignment requests from other IaaSs when needed. On the other hand, the *Composition Engine* orchestrates a set of elementary robots services in order to get the final required one when needed. The involved robots services are selected by the *Discovery Engine* (action 9). The latter selects the descriptors of available robots services for a given a request. It runs on the local repository to get the local services descriptors (action 10.1) and on the remote marketplace to get the available robots services descriptors from the remote marketplace (action 10.2). The *Publication* and the *Discovery Engine*s are presence technology-based (see Section V.B). Finally, IaaS monitors the robots to better schedule the workload and optimize the tasks delegation. The *Robot Monitor* is the module responsible for monitoring the origin robots in the *Physical Resources Layer* (action 11). A robot basically sends a notification to this module when it finishes its sub-task or fails. Accordingly, the robots availability is updated in the local *Robots Repository* (action 12) and in the external *Robots Services Marketplace*.

### C. Interfaces

The interfaces are designed according to the REpresentational State Transfer (REST) principle [16]. They all expose CRUD operations. For instance, the *Virtual Robots Management Interface* is a management interface that allows administrators to add/remove robots to IaaS. The *Publication Interface* allows IaaS to (un)publish its robots in the remote marketplace. The *Virtual Robots Management Interface* is the main interface. It exposes to the PaaS control operations to request robots services from IaaS. Table I details the list of the *Publication Interface* operations (the interfaces defined in [17] are re-used and modified according to the proposed architecture). These operations allow the *Publication Engine* to publish and update the presence information of its robots and allow the *Discovery Engine* to (un)subscribe to the presence information of robots belonging to other IaaSs.

Finally, it should be noted that the proposed IaaS for robots reuses and adapts the regular control and signaling IaaS interfaces. The interface between the network- and the node-level virtualization layers is one example. *Robot Monitor* and *Task Delegator* modules interact with the local robot through this interface. Its detailed specification is presented in [8].

## V. IaaS for Robots: Features And Technologies

This section discusses the features and used technologies of the proposed architecture, which contribute flexibility and cost efficiency in handling the robots. The goal is to validate the design presented in Section IV and to prove that it meets the architectural requirements discussed in Section II.

### A. Unified Description Model for Robots

The considered physical robots have different and various characteristics (e.g., capabilities, shapes). To meet the first requirement, a common model that unifies the robotic characteristics description is designed. The relevant literature considers developing ontologies for the standard description of heterogeneous resources (e.g., [18] for the specific case of robots). Although the semantics enable powerful and faithful modeling, its overhead in terms of processing and developing and maintaining new components is important. The proposed model extends the IETF Sensor Markup Language[1] (SenML) standard. SenML defines media types for representing simple sensor measurements and device parameters. It is lightweight and can be parsed efficiently, which makes it more suitable for the robots description.

TABLE II. STATIC CHARACTERISTICS REPRESENTATION USING EXTENDED SENML

| Static Characterist | SenML | JSON | Type |
|---|---|---|---|
| | Physical Char. | ph | Array |
| | Sensors | sen | Array |
| | Actuators | act | Array |
| | Personal Info. | info | Array |

| Sensors | SenML | JSON | Type | e.g. |
|---|---|---|---|---|
| | Sensor Name | sname | string | Camera, microphone |
| | Sensing Value Range | sval | string | (min, max) |
| | Sensing Unit | su | string | Hz for microphone |

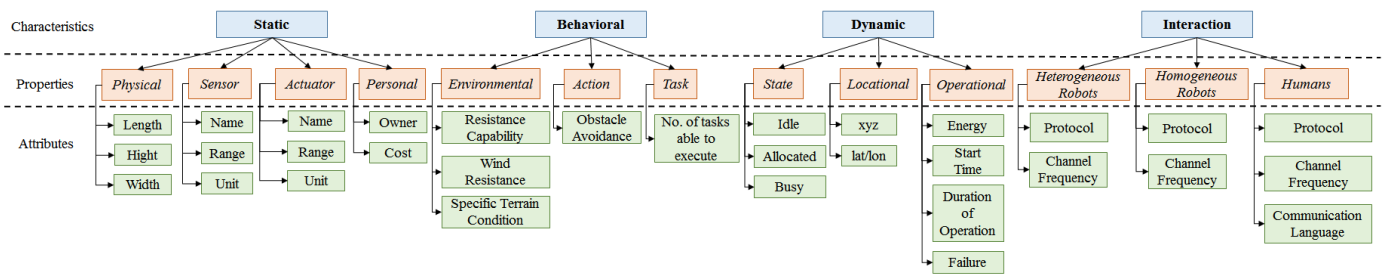

*Fig. 3. Extended SenML for Unified Robots Description Model*

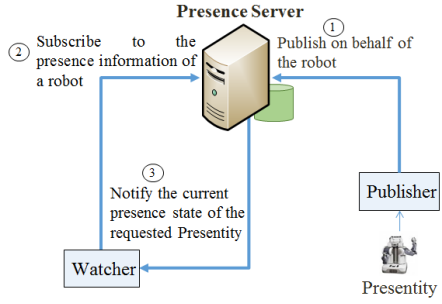

Fig. 4. Presence Technology-based Architecture for Publication/Discovery

The unified description model is implemented in the *IaaS O&M Manager*. It is this module that generates the generic robots descriptor to be stored in the local *Robots Repository* and in the remote *Robots Service Marketplace*. The robots characteristics are categorized into static, behavioral, dynamic, and interaction characteristics. Each characteristic includes a list of properties and each property may include one or several attributes. Fig. 3 shows the scheme of this model, with some examples of attributes for each property. Table II, for instance, details the properties and the Sensor attributes of the static characteristic.

### B. Publication and Discovery of Robots Services

The publication/discovery mechanisms are based on the presence technology [19]. These mechanisms enable meeting the second requirement. The presence service is chosen as it allows the discovery to be speedy and ahead of time. It allows each IaaS to publish its local robots whenever they change their state. This guarantees that the relevant robots are already discovered when IaaS receives a new search and rescue task. In the proposed architecture, the presence server is provided by a third-party tier (i.e., the *Robots Services Marketplace*). The *Publication and Discovery Engines*, as part of IaaS, are the clients that interact with the presence server. Fig. 4 shows a presence-based architecture for publication and discovery. The *Presentity* represents a robot. It is the source of the presence information to be stored and distributed by the presence server. The *Watcher* represents the *Discovery Engine*. It subscribes to a *Presentity* (i.e., robot) to receive its presence information along with its characteristics from the *Presence Server*. The *Publisher* represents the *Publication Engine*. It publishes the robots' presence information along with their characteristics on behalf of the robots in the *Presence Server*. It uses the SenML-based descriptors stored in the *Robots Repository*. This design allows each IaaS to publish its local robots when they change their state.

### C. Robots Services Composition

In a given IaaS for robots, the local and remote robots services can be orchestrated together to deliver complex services for the search and rescue applications. This enables meeting the third architectural requirement. The designed orchestration technique is simplistic. It consists of merging the related SenML-based descriptors of the considered robots services. This is done by the *Composition Engine*. The *Virtualization Engine* will use the resulting descriptor as a pattern to accordingly form the complex coalitions of robots, based on the composited SenML descriptors.

### D. IaaS for Robots Federation

The unified description model for the robots as well as the universal *Robots Service Marketplace* enable meeting the fourth requirement concerning the federation of several IaaSs for robots. All the involved IaaSs use the very same publication and discovery procedures when provisioning robots services. The whole available services are indexed, they are available in the centralized marketplace and are handled through the presence server. Furthermore, the common control and signaling interfaces enable the interoperability between these infrastructures. These interfaces are designed based on the REST principle to support the establishment and management of federation agreement. For instance, a *Request Handler* can receive and process a request coming from another IaaS as part of cooperation scenario. Similarly, during the runtime, a *Task Delegator* can schedule and assign tasks to remote robots that belong to a remote IaaS.

## VI. PROOF OF CONCEPT AND EXPERIMENTATIONS

This section describes the developed prototype that is implemented to validate and evaluate the findings of this work. It implements a search and rescue application that operates in earthquake sites.

### A. Prototype Description

The prototype implements a fire suppression functionality. It represents a sub-task of an earthquake search and rescue application. Earthquakes are often followed by fire with devastating consequences especially in townscape environments (e.g., Kesennuma City in Miyagi, Japan, 2011). The communication between different domains in the prototype is done through an ad-hoc network. It is assumed that the ad-hoc network is already built. This is necessary for disaster scenarios, where the telecommunication infrastructure is most likely crashed by physical destruction or the congestion of the network [20]. This could block out the system and hence prevents its functioning.

The prototype architecture is depicted in Fig. 5. In the *Physical Robots Domain*, the considered robots are LEGO Mindstorms NXT[2]. Two types of robots are used: one with arms and a movement motor and another with light sensors, kicking arms, and a movement motor. They carry plastic balls as water extinguishers. In the *IaaS Domain*, four distinct infrastructures are implemented. Each supports two robots services. The inter-domain architectural modules of IaaS (e.g., *Request Handler*, *Task Delegator*) are implemented as RESTful Web services using Java Restlet framework. The rest of the modules (e.g., *Virtualization Engine)* are developed as regular Java tools. The local *Robots Repositories* are simple OS folders that store the SenML-based descriptors of the supported robots services. In the *Gateways Domain*, an appropriate *Robot Gateway* is settled to map between the IaaS HTTP Java REST and LeJOS NXJ Java API commands that implement the Lego Communication Protocol (LCP). The *Robots Services Marketplace Domain* provides the presence server. It is implemented as RESTful Web Service using Restlet framework. It also includes a storage folder of the SenML descriptors of the published robots services.

---
[1] tools.ietf.org/html/draft-ietf-core-senml-09
[2] lego.com/en-us/mindstorms

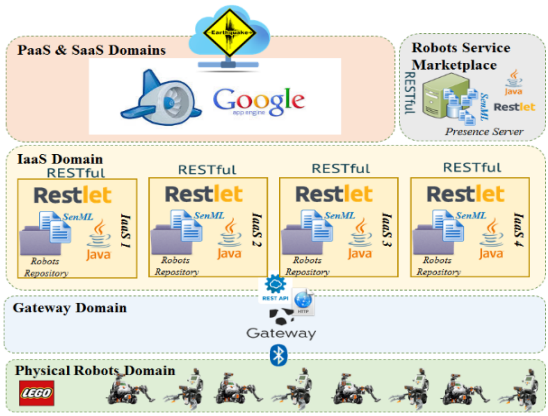

*Fig. 5. Prototype Architecture*

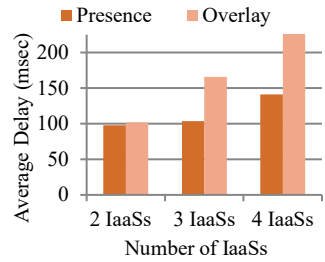

*Fig.6. Idle Robot Discovery Delay (IRDD)*

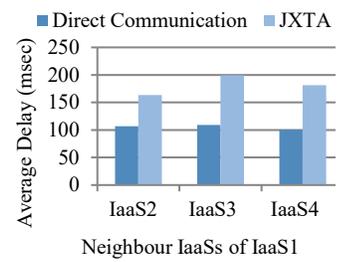

*Fig. 7. Task Assignment Delay (TAD)*

In the *PaaS Domain*, Google App Engine[3] (GAE) is used. It hosts and executes the search and rescue application with its fire suppression sub-task. Internet connection is assumed available between GAE and IaaS. A Network Address Translation (NAT) server is developed to redirect the requests coming from GAE to *IaaS*. In the *SaaS*, the fire suppression sub-task requires the light sensor capability to detect the balls and the kicking arms and movement capabilities to handle and move them. Consequently, the available robots services are composed before being bound to the application.

### B. Experimentation Setup

To properly evaluate it, the prototype is compared it with a peer-to-peer (P2P) overlay network. Consequently, an overlay node corresponding to each IaaS is implemented. An overlay node is implemented using the JXTA protocol (JXSE 2.6). The publication and the discovery procedures are carried out by the JXTA advertisement. The task assignment procedure is mapped to the JXTA messages that are exchanged through JXTA bidirectional pipes.

Four machines belonging to the same LAN are used, each hosts one IaaS for Robots. The first machine executes *IaaS1* and *NAT Server*. The second executes *IaaS2* and *Presence Server*. The third and the fourth respectively execute *IaaS3* and *IaaS4*. One of the machines has two interfaces: one with a public IP to communicate with the application and the other with a private IP (the LAN interface). The other machines have only a private IP. All machines run on Windows 7 Professional and have an Intel Core i7-2620 CPU with 2.70Hz and 8 GB of RAM.

### C. Measurements

The purpose of the performed experiments is the evaluate inter-IaaS communication, i.e., the federation cost including the publication/discovery procedures. To that end, two metrics are defined: 1) *Idle Robot Discovery Delay (IRDD)*, which is the time needed in milliseconds to get the requested *Presentities* state. This delay starts to be calculated as soon as IaaS subscribes to the marketplace presence information. 2) *Task Assignment Delay (TAD)*, which is the time difference in milliseconds between the moment IaaS sends a task assignment request and when the service's hosting IaaSs receive this request. We have already evaluated the virtualization and orchestration costs along with the system scalability in terms of the supported robots in our previous work [8].

*Test case 1 - Idle Robot Discovery Delay:* Fig. 6 shows the average time for IRDD using a various number of IaaSs. As is noticed, for any number of IaaSs, the average IRDD for presence-based publication/discovery is less than the delay for P2P overlay-based publication/discovery. This is because overlay networks have additional costs caused by the communication overhead. They add an intermediate level between the IaaSs. It is also observed that the average IRDD using P2P overlay increases as the number of IaaSs increase since IaaS1 should discover robots in more than one IaaS. The overlay nodes add additional overhead due to the processing of each packet. This shows the viability of using a presence technology-based publication/discovery

*Test case 2 - Task Assignment Delay:* Fig. 7 shows the average delay for TAD. In this test case, the average delay is calculated for the *Task Delegator* in the three IaaSs required to receive the task assignment request from the *Task Delegator* in IaaS1. It is observed that the average delay for direct communication remains almost the same for the three IaaSs. The involved IaaSs in the federated system communicate with each other directly, i.e., point-to-point. So, all IaaSs have the same delay. Moreover, this delay is far less compared to P2P overlay-based task assignment. To be received by each IaaS, the task assignment request needs to go through the overlay that adds overhead and increases latency in the system. This shows the viability of the proposed method for task assignment.

---
[3] appengine.google.com

## VII. CONCLUSION

This paper proposes cloud infrastructures for robots. Based on the designed business model, these IaaSs can be federated and cooperate to deliver robots *as-a-Service* to the cloud applications. The architecture along with its validating prototype are examined in a search and rescue application for a large-scale disasters case study. It meets all the identified requirements, such as a unified description model for the robots' capabilities and universal marketplace. The developed prototype shows the feasibility of this approach and evaluates the cost of federating several IaaSs.

Several lessons were learned during this research. The first one concerns the difficulty to homogenize the same node-level virtualization procedures for the whole IoT resources. Unlike VMs, where the underlying resources accord with the same fundamentals, there are fundamental differences when it comes to IoT resources such as robots and WSNs. For instance, it is obvious that two different applications cannot concurrently run two robots services provided by the same robots but they send contradictory movement actions. The same problem is not valid in the case of WSNs, when another set of specific characteristics, such as low computation and autonomy capabilities, need to be considered. These differences make it difficult to design one general IoT IaaS. Most of the existing works in the cloud are dedicated solutions for either WSNs or for robots.

The second lesson learned concerns the unsuitability of the classical technical services and governance for the newly integrated IoT services including robots. For instance, existing SLA and QoS management procedures in IaaS fail to handle robots services. This is due to the robots' characteristics, such as mobility. Unlike the classical IaaS resources (e.g., data centers), most of the considered physical robots are mobile, which considerably increases the probability of QoS degradation. Novel and sophisticated SLA and QoS management procedures need to be defined. Appropriate autonomic loops are among the prospective alternatives.